\documentclass[pre,reprint,floatfix,tightenlines,nofootinbib,superscriptaddress,showpacs]{revtex4-1}%
\pdfoutput=1

\usepackage{amsmath}
\usepackage{amssymb,amsfonts}
\usepackage{bm}
\usepackage[mathcal]{euscript}
\usepackage[pdftex]{graphicx}
\graphicspath{{figs/}}
\usepackage{subfigure}
\usepackage{color}
\usepackage{url}

\newcommand{\mathnotation}[2]{\newcommand{#1}{\ensuremath{#2}}}

\newcommand{\forget}[1]{}

\mathnotation{\vb}{\mathbf{v}}
\mathnotation{\htop}{h} 
\mathnotation{\e}{\mathrm{e}}
\mathnotation{\ldef}{\mathrel{\raisebox{.069ex}{:}\!\!=}}
\mathnotation{\rdef}{\mathrel{=\!\!\raisebox{.069ex}{:}}}
\mathnotation{\wb}{w_{\mathrm{B}}} 
\mathnotation{\map}{\mathcal{M}}

\mathnotation{\tnot}{t_0}
\mathnotation{\xf}{x_{\mathrm{f}}}
\mathnotation{\xe}{x_{\mathrm{e}}}
\mathnotation{\T}{T} 
\mathnotation{\C}{C} 
\mathnotation{\Cinf}{C_\infty} 

\mathnotation{\BM}{g} 
\mathnotation{\BMpar}{h} 
\mathnotation{\A}{a}
\mathnotation{\eps}{\epsilon}

\mathnotation{\visc}{\nu}
\mathnotation{\wu}{\mathcal{W}_{\mathrm{u}}}
\mathnotation{\ws}{\mathcal{W}_{\mathrm{s}}}
\mathnotation{\diff}{\kappa}
\mathnotation{\F}{f}
\mathnotation{\U}{U}
\mathnotation{\X}{x}
\mathnotation{\Y}{y}
\mathnotation{\xb}{\mathbf{x}}
\mathnotation{\ttau}{\tau}
\mathnotation{\R}{r}
\mathnotation{\N}{N}
\mathnotation{\D}{d}
\mathnotation{\la}{\Lambda}
\mathnotation{\Cself}{\tilde{C}}
\mathnotation{\Le}{L}
\mathnotation{\Ri}{R}
\mathnotation{\Ds}{Ds}
\mathnotation{\h}{h}
\mathnotation{\sigSE}{\sigma_{SE}}
\mathnotation{\CSE}{C_{SE}}
\mathnotation{\boxw}{\ell}
\mathnotation{\Acal}{\mathcal{A}}
\mathnotation{\sigC}{\sigma_C}
\mathnotation{\flambda}{f_\lambda}

\begin{document}

\title{Measures of mixing quality in open flows with chaotic
advection}
\author{E. Gouillart}
\affiliation{Surface du Verre et Interfaces, UMR 125 CNRS/Saint-Gobain, 93303
Aubervilliers, France}
\author{O. Dauchot}
\affiliation{Service de Physique de l'Etat Condens\'e, DSM, CEA Saclay,
URA2464, 91191 Gif-sur-Yvette Cedex, France}
\author{J.-L. Thiffeault}
\affiliation{Department of Mathematics, University of Wisconsin --
  Madison, WI 53706, USA}

\date{\today}


\begin{abstract} 

  We address the evaluation of mixing efficiency in experiments of
  chaotic mixing inside an open-flow channel. Since the open flow
  continuously brings new fluid into the limited mixing region, it is
  difficult to define relevant mixing indices, as fluid particles
  experience typically very different stretching and mixing histories.
  The repeated stretching and folding of a spot of dye leads to a
  persistent pattern. We propose that the normalized standard
  deviation of this characteristic pattern is a good measure of the
  mixing quality of the flow.  We discuss the link between this
  measure and mixing of continuously-injected dye, and investigate it
  using an idealized map.

\end{abstract}
\vspace{-0.5cm}

\maketitle

\section{Introduction}

Mixing viscous fluids (whether Newtonian or not) in throughflow
reactors is one of the most common and important industrial
process~\cite{handbook}.
Relevant quantitative measures of
mixing efficiency are crucial for the design and evaluation of mixing
systems. However, because of the wide spectrum of applications, there
is not to date a universally accepted way of defining mixing
efficiency~\cite{Ottino1989}.

Since the 1980s, mixing at low Reynolds number has been analyzed by
fluid dynamicists in the framework of chaotic
advection~\cite{Aref1984, Ottino1989}, that is, the capacity
of a mixing device to generate Lagrangian trajectories that separate
exponentially fast. In closed-flow mixing, that is when inhomogeneous
fluid contained in a closed domain is stirred, some mixing indices
stem directly from an understanding of the mechanisms of chaotic
advection~\cite{Ottino1989}. Among classical measures, Poincar\'e
sections represent the region explored by a trajectory stroboscoped at
discrete timesteps, and hence give the extent of the chaotic
region~\cite{Leong1989, Ottino1990, Jana1994} for time-periodic
systems. Lyapunov exponents account for the stretching experienced by
fluid particles~\cite{Muzzio1991, Muzzio1992, Antonsen1996}.
Topological entropy~\cite{Boyland2000, Stremler2007, Gouillart2006,
  Finn2006} and braiding factors~\cite{Thiffeault2005, Thiffeault2010}
describe the entanglement of Lagrangian trajectories.  Such
quantities, however, are hard to link directly to the degree of
homogeneity achieved by the concentration field of a scalar to be
mixed. More direct indices of homogeneity include the decaying
variance of the scalar field~\cite{Williams1997, Jullien2000,
  Burghelea2004, Villermaux2003, Duplat2008} or its
entropy~\cite{Wang2003}. In some periodic flow configurations, the
concentration field converges rapidly to a persistent pattern that
repeats over time with decaying contrast~\cite{Pierrehumbert1994,
  Rothstein1999, Voth2003, Liu2004, Tsang2005}.  Such a pattern is a
specific eigenmode of the Floquet operator of the advection-diffusion
equation, dubbed \emph{strange eigenmode}; as such it allows an
intrinsic characterization of the mixing flow.

\begin{figure}
\centerline{\includegraphics[width=0.98\columnwidth]{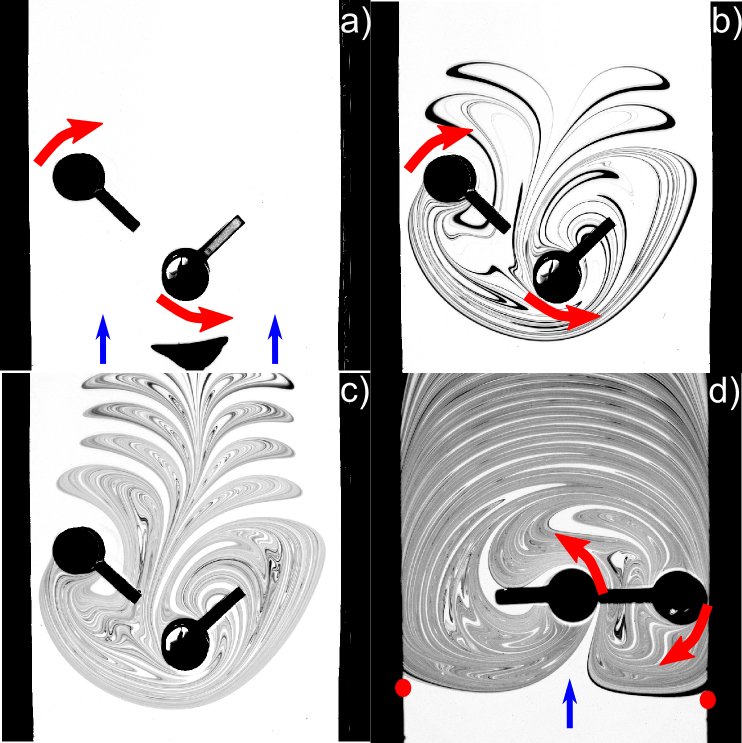}}
\caption{(a) Experimental setup: a viscous fluid in a channel is stirred
  by two rods moving in an eggbeater motion.  A spot of black dye is
injected upstream of the rods. (b) Mixing pattern observed shortly after
the dye has entered the vicinity of the rods: the rods stretch some dye
particles into thin filaments, while others escape downstream without
being stretched and mixed. (c) Mixing pattern observed at longer times:
the intensity of dye in the pattern decays with time due to diffusion --
this is the `strange eigenmode' regime. (d) If the rods are rotated in
the opposite direction to (a--c), they prevent fluid from flowing along
the channel walls. Two stagnation points (plotted as disks) on the
channel sides trap fluid in their vicinity for long times. In that case,
the mixing pattern occupies the whole width of the channel.
\label{fig:mixing_sys}}
\end{figure}

Less studied is the case of chaotic advection in open flows, in which
fluid enters and then exits a limited mixing region. In this region,
mechanical stirrers lead to exponential separation of fluid particles
for the duration of their stay in the region (see
Fig.~\ref{fig:mixing_sys}(a--b)).  Extensive studies of idealized open
flows in 2D have focused on the properties of the underlying dynamical
systems, such as the multifractal set of never-escaping trajectories
-- called the \emph{chaotic saddle} -- and its influence on Lagrangian
trajectories flowing through the mixing
region~\cite{Jung1993,Pentek1995,Sommerer1996,Neufeld1998,Pentek1999,Tel2000,
  Tel2005}.  Such studies focused mostly on reaction rates, and less
on the mixing quality.

On the other hand, a variety of practical indices of mixing quality
for open reactors can be found in the chemical engineering literature
\cite{Danckwerts1952, Bryant1977, Ehrfeld1999, Aubin2003,
  Kukukova2009}.  These indices are mostly used for turbulent
flows. The \emph{intensity of segregation} was defined by
Danckwerts~\cite{Danckwerts1952} as the variance of the scalar field
in the outflow. This index characterizes directly the homogeneity,
hence the quality, of the end product. Refined measures of mixing take
into account the spatial scale of the scalar field as well as its
variance~\cite{Kukukova2009}. One can also compute the departure of
the residence-time distribution (RTD) of fluid particles inside the
mixer from a purely exponential RTD that characterizes a \emph{perfect
  mixer}, in the language of chemical engineering.

In recent work~\cite{Gouillart2009a}, we have studied the mixing
dynamics of quasi-2D free-surface channel flows stirred by two rods
moving in an eggbeater protocol (see Fig.~\ref{fig:mixing_sys}(a) for
a picture of the rods and the channel). A spot of dye released
upstream of the stirrers (Fig.~\ref{fig:mixing_sys}(a)) is advected by
the main channel flow and enters the mixing region where it is
stretched and folded by the stirrers. Some dye particles leave the
mixing region very quickly (Fig.~\ref{fig:mixing_sys}(b)), while other
parts of the spot remain in the mixing region longer
(Fig.~\ref{fig:mixing_sys}(c)). We demonstrated that, after a short
time, the concentration field of the dye takes on a persistent pattern
whose mean intensity decays with time~\cite{Gouillart2009a}. By
analogy with persistent patterns in closed
flows~\cite{Pierrehumbert1994, Rothstein1999, Voth2003}, we called
such patterns \emph{open-flow strange eigenmodes}.

In the present work, we examine which information about mixing
quality can be gleaned from such experiments in open flows. In
particular, we investigate whether relevant measures of mixing
quality can be derived from the concentration field of the strange
eigenmode. This pattern is very robust, in the sense that it does not
depend on the upstream location where dye was injected. It is
therefore natural to use the pattern to characterize the mixing
device.

The paper is organized as follows. We first describe the experimental
apparatus and methods in Section~\ref{sec:device}. In
Section~\ref{sec:history}, we describe the history of a spot of dye
passing through the mixer, and the evolution of the resulting
concentration field.  This helps to extract relevant information from the
concentration images and time series, as we describe in
Section~\ref{sec:mixing_infos}.  In particular, we introduce a new mixing
index called the \emph{eigenmode index}, which is the normalized standard
deviation of the persistent concentration pattern.  In
Section~\ref{sec:Is}, we relate this index to the intensity of
segregation for a continuous injection of heterogeneity as introduced by
Danckwerts~\cite{Danckwerts1952}. The paper ends with a summary and
discussion of the results.

\section{Mixing device configurations
\label{sec:device}}

Experimental data are obtained in an open-flow channel whose
experimental set-up has been described in more details
in~\cite{Gouillart2009a, Gouillart_thesis}. Viscous fluid (cane sugar
syrup) flows at a fixed rate through a long and shallow transparent
channel. Two rod-stirrers (see Fig.~\ref{fig:mixing_sys}) travel on
intersecting circular paths at mid length of the channel, in an
eggbeater protocol. Because of the low Reynolds number (see below),
the mixing flow is time-periodic. We select the eggbeater protocol for
the simplicity of its geometry as well as its ability to promote
efficient chaotic advection. Indeed, the intersection of the
trajectories of the rods ensures chaotic advection, due to stretching
and folding of fluids particles. Through the stirring frequency, we
control the order of magnitude of the mean number of periods spent by
fluid particles inside the mixing region, which was varied between $4$
and $24$ in our experiments~\cite{Gouillart2009a}. Depending on the
stirring frequency, Reynolds numbers range from 2 to 10, well within
the laminar regime. The mixing region may be defined loosely by the
region spanned by the trajectories of the stirrers; a more rigorous
definition will be given in the next section.

The orientation of the trajectories of the rods (with respect to the
direction of the main flow) determines the routes along which fluid
flows preferentially inside the mixing region. We may rotate the rods
so that they assist fluid in passing along the channel walls
(Fig.~\ref{fig:mixing_sys}(a)--(c)), where they drag fluid inside the
mixing region only while on the upper and central part of their
trajectory.  Or we can rotate the rods in the opposite direction, so
that fluid moving along the side walls cannot cross the mixing region,
but is forced into a central funnel (Fig.~\ref{fig:mixing_sys}(d)). We
call the two cases respectively protocol type A and protocol type B.

We perform mixing experiments by releasing a spot of dye upstream of
the rods (Fig.~\ref{fig:mixing_sys}(a)). Pictures of the evolving dye
pattern are taken at every period of the stirring motion with a
high-resolution digital camera. Our set-up has been designed
to measure quantitatively the concentration field from image
processing (see~\cite{Gouillart2009a, Gouillart_thesis} for more
details). This requires in particular using only backlighting of the
dye pattern.

\section{History of a dye spot \label{sec:history}}

We now turn to a qualitative description of the different spacetime
trajectories of fluid particles, and of the resulting dye concentration
fields. The understanding we gain here will be used in the next section
for defining mixing indices.

From the moment when the spot of dye is injected upstream of the rods,
the fluid particles in the spot can have two types of radically
different future histories: they either enter the stirring region and
are mixed to some degree, or they pass through the region without
being mixed.  The latter behavior occurs predominantly in type A
protocols, where fluid particles can flow freely along the channel
walls.

The patch of fluid particles that enter the mixing region at a given
period consists of dyed particles as well as some undyed white
particles.  After a few stirring periods, only white undyed fluid
enters the mixing region.  Before diffusion becomes important, the
patch containing dyed particles is stretched and folded by the rods.
At each stirring period, the patch moves to a new part of the mixing
region, and its previous location is replaced by fresh, undyed fluid
that entered the region later.

As time goes on, the original dyed patch inside the mixing region
develops thin filaments as it is stretched and folded by the rods (see
the filaments inside the mixing region in
Fig.~\ref{fig:mixing_sys}(b)). Nevertheless, as long as the width of
the filaments is large enough, fluid particles of the dyed patch are
not mixed by diffusion with undyed fluid particles. For example, in
Fig.~\ref{fig:mixing_sys}(b) most dye filaments bear the same
concentration level as the initial dye spot. At every stirring period,
some particles of the dyed patch leave the mixing region to compensate
for fluid entering the mixing region; for short residence time they
have not been mixed at all with undyed particles (see the dark
filaments in the upper part of Fig.~\ref{fig:mixing_sys}(b),
downstream of the rods).

For longer residence times, mixing by diffusion of fluid particles with
different entry times occurs at locations where the patch has been
compressed down to the diffusion scale at which stretching and diffusion
balance -- the Batchelor scale~\cite{Batchelor1959,Villermaux2003,
Gouillart2008a}. The Batchelor scale is given by
\begin{equation}
\wb = \sqrt{\diff/\la}, 
\end{equation}
where $\diff$ is the diffusion coefficient and $\la$ is the mean
stretching (Lyapunov exponent). The scale $\wb$ is the smallest that
can be observed in the mixing pattern, and increases with
$\diff$. Once the patch has reached the scale $\wb$, diffusion
efficiently blurs its particles with particles that have different
entry times. In particular, dye filaments are blurred with surrounding
white fluid, and intermediate gray concentration levels appear as in
Fig.~\ref{fig:mixing_sys}(c).  Mixing is efficient for particles of
the patch that have stayed inside the mixing region at least until
their filaments reach $\wb$.

Such filaments approach the (non-elliptic) \emph{periodic points} of
the flow, which return to the same position after a given number of
stirring periods and hence remain forever inside the mixing
region. The set of all periodic points is known as the \emph{chaotic
  saddle} for open flows~\cite{Tel2000, Tel2005, Pentek1995}. It is a
set of zero measure.  The periodic points are typically located in
between the regions occupied by the patch for short residence times,
which cannot contain periodic points as they are filled with new fluid
at each period. The mixing region can now be defined more precisely as
the basin where the orbits of the chaotic saddle remain trapped
forever, i.e., the convex hull of the set containing all these periodic
orbits. Using arguments from lobe dynamics~\cite{Beigie1991}, it can
be shown that this set is delineated on the upstream side by the
upstream boundary of the strange eigenmode pattern. On the downstream
side, the extent of the mixing region is bounded by the symmetric line
of the upstream boundary with respect to the horizontal axis linking
the rotation center of the rods. 

Dye filaments approach a periodic orbit of the chaotic saddle along the
orbit's \emph{stable manifold} -- the set of fluid particles that
converge to the periodic orbit. Conversely, they escape an orbit along
its \emph{unstable manifold} -- the set of particles that diverge from
the orbit.  The typical situation is that fluid particles in the patch
approach an orbit of the chaotic saddle along its stable manifold, then
recede along its unstable manifold towards another orbit, and so on until
they finally escape downstream.

The dye pattern of Fig.~\ref{fig:mixing_sys}(c) traces the unstable
manifold of the chaotic saddle. As the unstable manifold is invariant
under a full stirring period, dye always returns to the same
pattern. We observe indeed in our experiments that patterns such as
the one in Fig.~\ref{fig:mixing_sys}(c) are persistent in time.
Because of diffusion and the loss of dye downstream, the intensity of
the pattern becomes weaker and weaker.

\begin{figure}
\begin{minipage}{0.32\columnwidth}
\centerline{\includegraphics[width=0.9\columnwidth]{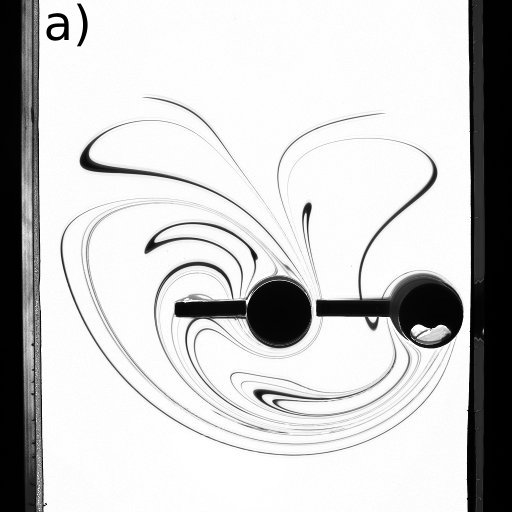}}
\end{minipage}
\begin{minipage}{0.32\columnwidth}
\centerline{\includegraphics[width=0.9\columnwidth]{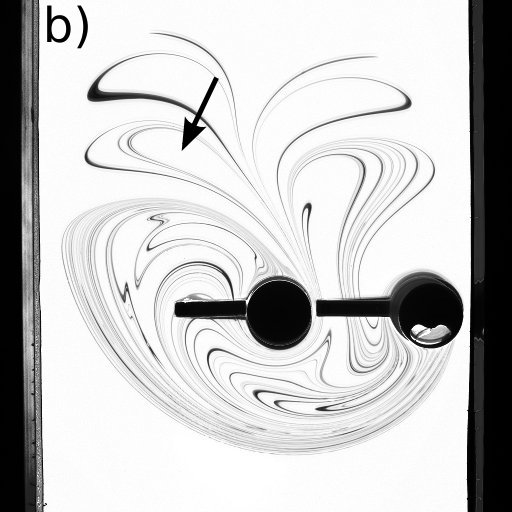}}
\end{minipage}
\begin{minipage}{0.32\columnwidth}
\centerline{\includegraphics[width=0.9\columnwidth]{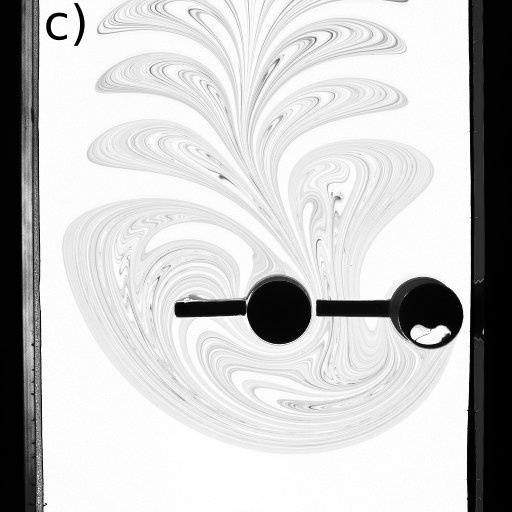}}
\end{minipage}
\centerline{\includegraphics[width=0.9\columnwidth]{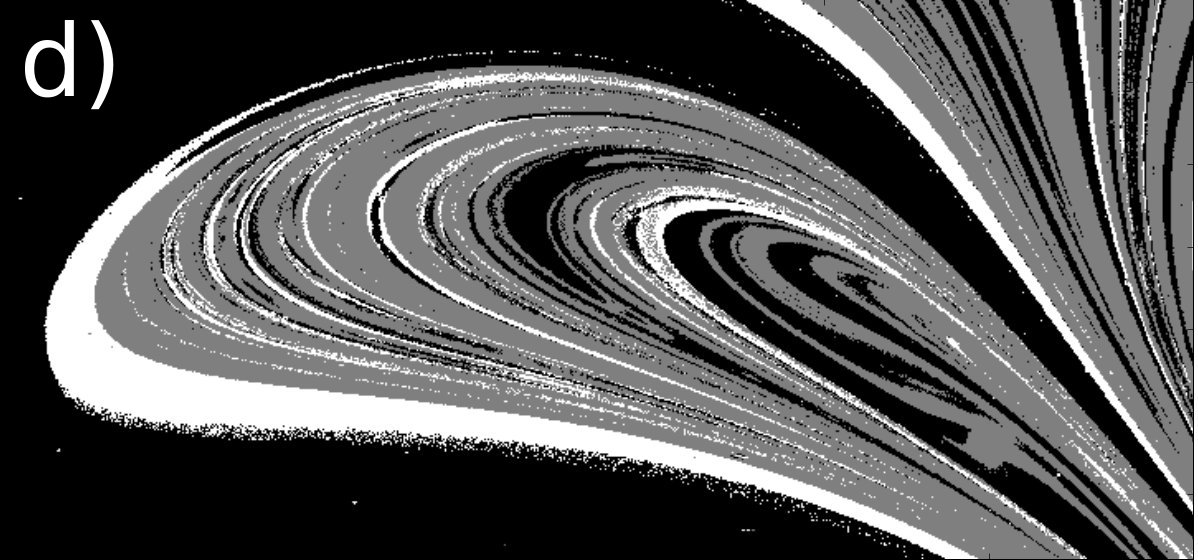}}
\caption{(a--c) Mixing pattern at periods 2, 3 and 6. Patterns~(a)
and~(b) correspond to times shortly after the dye spot has entered the
mixing region, while in (c) the dye pattern traces out the strange
eigenmode. Since the flow is deterministic, the supports of the
escaping lobes for the three different times are non-intersecting
domains. (d) For the same experiment as (a--c), blowup of a downstream
lobe, where we have superimposed the pattern of (a) and (b) (white), and
the long-time pattern of (c) (gray). Filaments with short residence times
are found in white `holes' of the long-time pattern.
\label{fig:scenar}}
\end{figure}

In previous work~\cite{Gouillart2009a}, we have shown that for protocol
type A the dye pattern converges to a persistent concentration field that
repeats itself periodically and whose mean is decaying exponentially with
time. The concentration field obeys the equation
\begin{equation*} 
\C(\xb, t=n\T) = \exp(-n\T/\tau)\, \Cself(\xb)
\end{equation*}
where~$\tau$ is the decay time ($1/\tau$ is the decay rate).  (In
general~$1/\tau$ is complex, which leads to oscillations around a
decaying trend; here we take its real part.)  By analogy with closed
flows~\cite{Pierrehumbert1994}, we call $\Cself(\xb)$ the
\emph{strange eigenmode} of the flow, as it is an eigenmode of the
advection-diffusion
operator~\cite{Gouillart2009a}. Fig.~\ref{fig:mixing_sys}(c) displays
an example of strange eigenmode. A physical picture of the support of
the strange eigenmode (the space where $|\Cself(\xb)| >0$) is given by
the unstable manifold of the chaotic saddle -- a fractal set of zero
measure~\cite{Tel2000} -- fattened by diffusion into filaments of
width~$\wb$. Conversely, the complement of the support of the
eigenmode does not contain any periodic orbits and corresponds to the
locations occupied by a patch of fluid shortly after its entry time,
before it has reached the Batchelor scale. It can be seen in
Fig.~\ref{fig:scenar} that the short-time iterates of the dye spot are
located inside the holes of the eigenmode. For example, the
filamentary pattern in the mixing region in Fig.~\ref{fig:scenar}(a)
is located in the holes of the long-time pattern of
Fig.~\ref{fig:scenar}(c). This is also true for the fluid escaping the
mixing region: the escaping filaments in the lobes of
Figs.~\ref{fig:scenar}(a--b) are found within the holes of the pattern
that periodically leaves the mixing region at long times
(Fig.~\ref{fig:scenar}(c)). The properties of the strange eigenmode
were investigated more thoroughly in~\cite{Gouillart2009a}.

For protocol type B, however, we found that the pattern continues to
evolve slightly with time and its decay does not exhibit true exponential
dynamics. The reason is that, in contrast to protocol type A, the mixing
pattern extends to the side walls of the channel (as in
Fig.~\ref{fig:mixing_sys}(d)) where the flow is less intense and
stretching is slower. Therefore, dye may remain for long times in the
vicinity of the walls, whereas it escapes much faster when far from the
walls. As a result, the dye is increasingly located in the vicinity of
the walls, and no persistent pattern is observed. Protocol type B is
characterized by two specific stagnation points, whose unstable manifolds
divide the upstream flow and the mixing region (red points in
Fig.~\ref{fig:mixing_sys}(d)). These points belong to the set of periodic
orbits, but fluid approaches them much more slowly (algebraically rather
than exponentially) than other periodic points. Fluid particles are
therefore exchanged more rapidly between the periodic orbits of the bulk
than between the wall region and the bulk.
(see~\cite{Gouillart2009a} for more details). This prevents the onset of
a persistent strange eigenmode.

In this article, we will mostly focus on the study of type A
protocols, where the dye pattern converges to strange eigenmode, but
we also address the case of type B protocols.

\section{Mixing measures derived from the evolution of a dye spot
\label{sec:mixing_infos}}

We now consider the different information about mixing efficiency that
can be gained from our experiments, particularly from the strange
eigenmode pattern.

\subsection{Location of well-mixed particles}

\begin{figure}
\begin{minipage}{0.48\columnwidth}
\subfigure[]{
\centerline{\includegraphics[width=0.9\columnwidth]{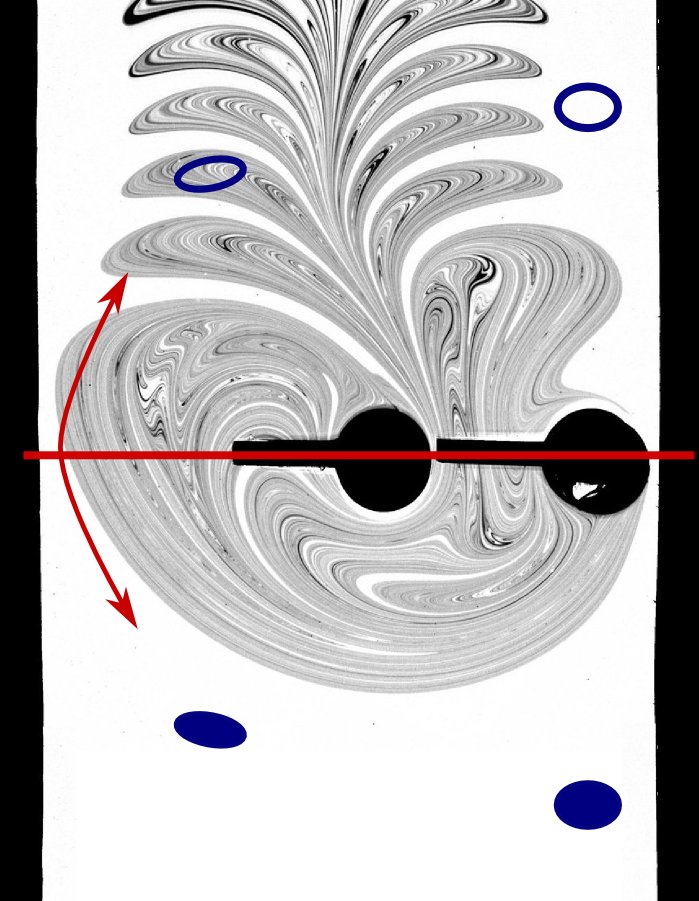}}
}
\end{minipage}
\begin{minipage}{0.48\columnwidth}
\subfigure[]{
\centerline{\includegraphics[width=0.9\columnwidth]{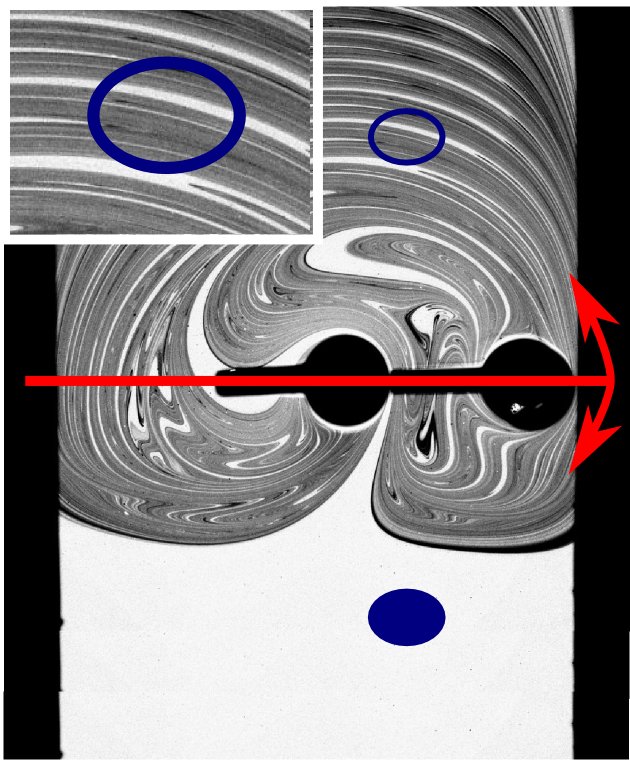}}
}
\end{minipage}
\caption{The fate of dye spots injected upstream (solid dark spots
sketched on (a) and (b)) can be inferred from the position of their
mirror image across the axis of the rods (hollow spots). (a) Example of a
type A protocol. Mirror images that intersect the strange eigenmode, such
as the left spot, correspond to well-mixed fluid particles. The right
spot, however, has no intersection with the strange eigenmode, and will
escape downstream without being caught by the rods and without being
mixed. (b) Example of a type B protocol. The dye spot falls on both the
support of the permanent pattern (gray patches) and its complement (white
stripes), meaning that only the portion located on the support of the
permanent pattern will be smeared by diffusion. It is therefore possible
to obtain detailed information, such as the fraction of a dye
spot that will be well-mixed, by superimposing the mirror image of the
spot onto the strange eigenmode pattern.
\label{fig:flip}}
\end{figure}

The eigenmode pattern can be used to give \emph{predictions} about the
efficiency of mixing for fluid particles injected at different locations.
We illustrate this with the example of two dye spots represented in
Fig.~\ref{fig:flip}(a), for which one would like to know if they will be
efficiently mixed by the rods. The degree of mixing experienced by a
fluid particle depends on its distance to the stable manifold
$\mathcal{W}_\text{s}$ of the chaotic saddle. Particles lying within a
distance $\wb$ of the stable manifold, where~$\wb$ is the Batchelor
scale, are very well mixed since, at long times, they lie within a
distance $\wb$ to the unstable manifold $\mathcal{W}_\text{u}$.  In other
words, they are eventually compressed onto the eigenmode pattern. On the
other hand, particles lying further from $\mathcal{W}_\text{s}$ escape
the region having experienced little mixing.

For our stirring configurations, an approximation to the stable
manifold $\mathcal{W}_\text{s}$ is easily obtained: it is the mirror
image (with respect to the axis of the rods' centers) of the strange
eigenmode pattern, which traces out the unstable manifold. This comes
directly from (i) the time-reversibility of Stokes flows, and (ii) the
symmetry of the trajectories of the rods with respect to the
horizontal axis linking their rotation centers. Mirror images are
therefore to be taken when both rods lie on the symmetry axis. It is
then possible to predict the fate of a fluid particle from the
intersection of its mirror image (blue circles in
Fig.~\ref{fig:flip}(a)) with the strange eigenmode. The intersecting
part of the spot will be well mixed, as opposed to the
non-intersecting part. In Fig.~\ref{fig:flip}(a), the dye spot on the
left is very well mixed as its mirror image is nested within a
densely-striated lobe of the eigenmode. The dye spot on the right,
however, escapes downstream without being mixed, as its mirror image
has no intersection with the strange eigenmode. This has been checked
experimentally.

In the same way, in Fig.~\ref{fig:flip}(b), the parts of the dye spot
whose mirror image is located on the strange eigenmode will be very
well mixed with surrounding fluid.  The other parts located on the
mirror image of white filaments are less mixed -- although they may
experience some stretching -- and will typically retain their initial
upstream concentration level once advected downstream of the rods.
Such considerations are useful to determine the ideal location to
inject a substance that is to be mixed. This ideal location, as shown
above, is \emph{inside the mirror image} of the strange eigenmode.

\subsection{Distribution of residence times}

\begin{figure}
\subfigure[]{
\centerline{\includegraphics[width=0.6\columnwidth]{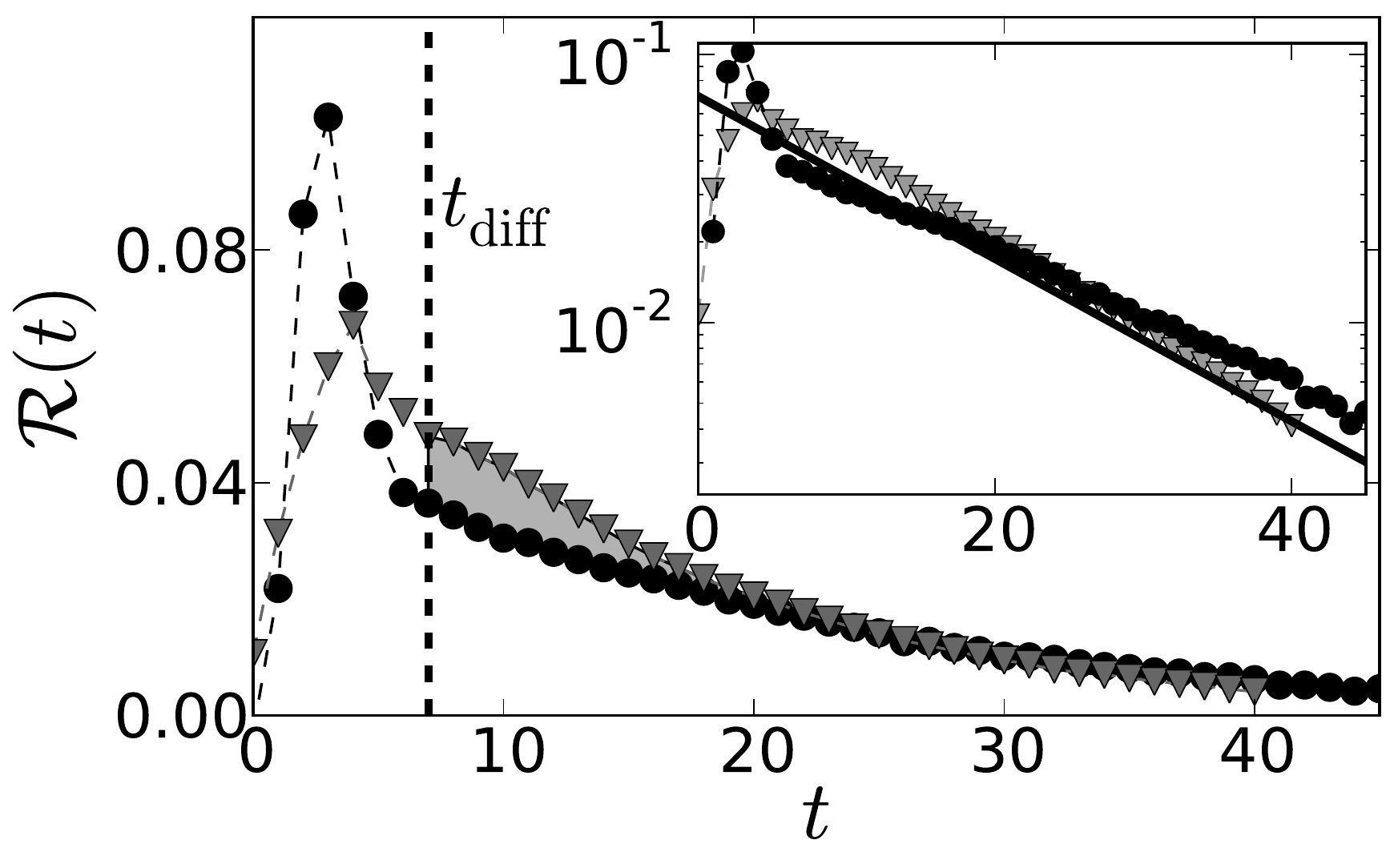}}
}
\subfigure[]{
\centerline{\includegraphics[width=0.6\columnwidth]{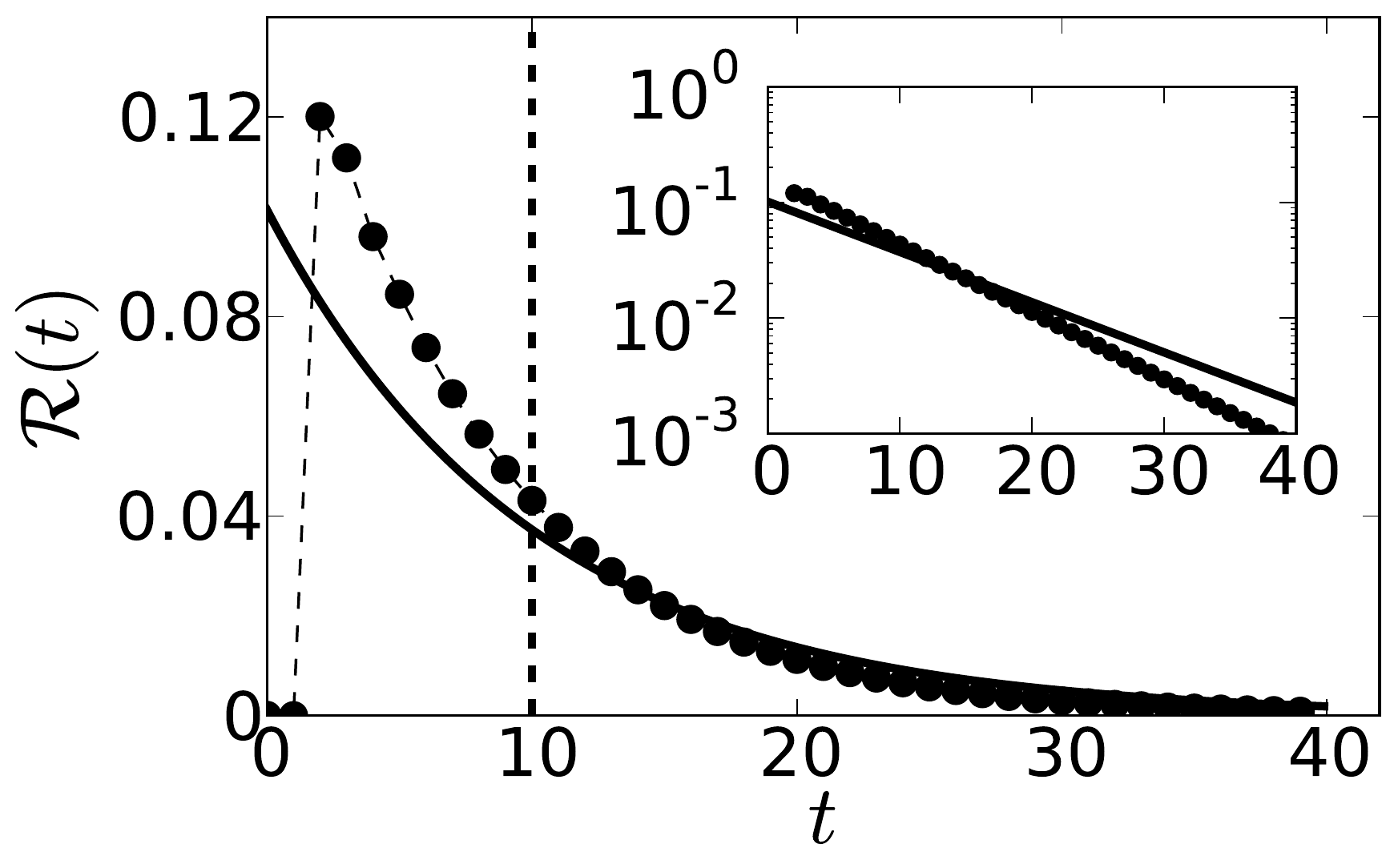}}
}
\caption{(a) \textbf{Experiments:} Residence time distribution (RTD)
  inside the mixing region for a typical A protocol (black
circles) and B protocol (gray triangles).  A vertical dashed line
represents $t_\text{diff}$, the time where diffusion starts blurring
fluid particles with different concentrations. The type A protocol is
less effective at mixing than the B protocol, as reflected by the
pronounced short-time peak of the RTD.  Inset: RTDs on a semi-log scale
display exponential tails, with slightly different slopes for each
protocol type.  The solid line shows the purely-exponential RTD of a
perfectly random mixer. (b) \textbf{Open-flow baker's map:}
Circles: RTD inside the mixing region (inset: RTD on a semi-log scale).
The minimum exit time is three periods. Solid line: exponential RTD of a
perfectly random mixer with the same flow rate. \label{fig:rtds}}
\end{figure}

We now consider another quantity of interest for mixing: the
residence-time distribution (RTD). Such a characterization of mixing is
very common in chemical engineering and was described in detail by
Danckwerts~\cite{Danckwerts1952}.  Since the residence time of fluid
particles inside the mixing region is related to the transport properties
of the flow, and, to some extent, to its mixing properties, the analysis
of the RTD can be used to obtain qualitative information about mixing. We
show below that an approximate quantitative measure of mixing quality can
also be derived from the RTD by distinguishing between residence times
that are greater or smaller than the typical time needed for diffusion to be
effective.

In our experimental device, we can determine the RTD for dye particles
coming from the initial dye spot by measuring the fraction of the dye
that escapes the mixing region during the time interval from~$t$
to~$t+T$. Generally, the residence time distribution can be measured
by recording the amount of dye at the outlet of the mixing region, either
in model experiments or even in industrial settings. It is
usually more convenient to measure the RTD than to obtain pictures of the
strange eigenmode, because of obstructions or three-dimensional complexity.

In Fig.~\ref{fig:rtds}(a) we plot the RTD for a type A and B protocols
with the same rotation speed (only the rotation direction is different).
Let us give first a qualitive analyses of the two RTD's. Two distinct
regimes are observed: (i) a peak for short residence times, which is more
pronounced for the type A protocol; followed by (ii) an exponential decay
regime (see the inset for a semilog plot of the RTD). For the type A
protocol, most dye particles that contribute to the short-time peak are
in fact not entrained inside the chaotic mixing region; they flow almost
unperturbed along the sides of the channel or, to a lesser extent, are
entrained inside the mixing region and expelled downstream immediately.
For both protocols, the onset of the exponential regime is associated
with particles that are trapped in the mixing region for some time.
These particles are close to the chaotic saddle, and are thus affected
by their proximity to its periodic orbits~\cite{Tel2000,Gouillart2009a}
(see Section~\ref{sec:history}).  Diffusion plays no role in this regime
until particles have reached the residence time~$t_\text{diff}$, the time
at which dye filaments reach the Batchelor scale.  This marks the onset
of the strange eigenmode regime and of good mixing.

A first quantitative index of mixing is therefore the fraction of
fluid particles that escape only after they have been stretched enough
to be blurred with other particles by diffusion:
\begin{equation}
\alpha_\text{diff} = \int_{t_\text{diff}}^{\infty} \mathcal{R}(t)\,\text{d}t,
\end{equation}
with $\mathcal{R}$ the RTD. Using this index, we see in
Fig.~\ref{fig:rtds}(a) that the type B protocol is more efficient from
a mixing perspective than the type A protocol: in the latter case,
many fluid particles move along the sides the channel without being
mixed.

However, this index suffers from two drawbacks. First, due to local
disparities of the stretching rate, some dye filaments reach the Batchelor
scale and start diffusing earlier than others. Specifically, some fluid
particles may spend a long time near weakly unstable orbits and thus
retain their initial concentration level, while most particles are well
into the diffusive regime.  It is therefore difficult to define a single
diffusion time.  Our choice for $t_\text{diff}$ in
Fig.~\ref{fig:rtds}(a) is a pragmatic one: it is the period when
\emph{most} filaments start fading from pure black to intermediate
gray. The second drawback is that the RTD -- hence $\alpha_\text{diff}$ --
depends on the location of the injected dye spot.  More precisely, it
depends on the fraction of the spot intersecting the mirror image of
the strange eigenmode, as was discussed above. For type A protocols,
for example, a dye spot injected at the center of the channel yields a
higher mixing index than a dye spot injected close to the channel
walls.  This problem can be resolved by covering with dye a strip that
encloses all fluid particles that enter the mixing region during one
period, or by dyeing all incoming fluid from a chosen starting time,
as proposed in \cite{Danckwerts1952}. These solutions present
experimental difficulties. In the next subsection we introduce a more
refined index based on the strange eigenmode pattern that does not
suffer from the two limitations mentioned above. 

In chemical engineering, distributions of residence times are often
used to compute another mixing index, that is the \emph{departure from
  perfect mixing}~\cite{Danckwerts1952}. A perfect mixer is
defined~\cite{Danckwerts1952, handbook} as a device where an incoming
fluid particle has a constant escape probability with time. It is
therefore characterized by an exponential RTD, with escape timescale
$\tau_Q= Q/\mathcal{V}$, with $Q$ the flowrate and $\mathcal{V}$ the volume
of the mixing region, according to mass conservation. Mixers that generate
vigorous turbulence at high Reynolds number, for example, shuffle
fluid particles randomly and may be close to perfect mixers --
although the minimum exit time is always non-zero for a real system.
Departures from this random mixing may be interpreted as a bypass
where fluid is whisked away from the mixing zone, or dead zones where
fluid is trapped for long times.  One may therefore define a mixing
index that measures the difference between the observed and perfectly
random residence time distributions:
\begin{equation}
\Delta\mathcal{R} = \int_0^\infty \left| \mathcal{R}(t) -
{\tau_Q^{-1}}\exp(-t/\tau_Q) \right| \text{d}t. 
\end{equation}  
For type A protocols, for example, the main contribution to
$\Delta\mathcal{R}$ comes from fluid particles advected rapidly along
the channel sides, and for these protocols this measure correlates
well with $\alpha_\text{diff}$.

However, in the general case there is no direct relation between
$\Delta\mathcal{R}$ and the fraction of fluid particles blurred by
diffusion.  Using $\Delta\mathcal{R}$ as a measure of mixing, it is
even possible to find mixers that have a better mixing efficiency than
the ``perfect mixer.''  For example, this will occur if there is a
minimum residence time for fluid particles in the mixing region --
this minimum time being greater than the diffusion time
$t_\text{diff}$.  This is impossible in our experiments, since the
rods transport some fluid particles rapidly from the upstream side to
the downstream side of the mixing region, so that very small residence
times are unavoidable.  A mixer with more rods arranged in an
ingenious way may nevertheless increase the minimum residence time up
to $t_\text{diff}$.

In Fig.~\ref{fig:rtds}(b), we show an example of such an RTD obtained
with the open-flow baker's map, where fluid particles circulate during
at least three periods before they can exit the mixing region (the map
was introduced in~\cite{Gouillart2009a} and is described in
Section~\ref{sec:openbaker}; it is similar to an earlier map
in~\cite{Neufeld1998}).  As fluid particles are stretched during this
residence time, this reduces the minimum scale of heterogeneities that
exit the mixer, and if $t_\text{diff}$ is small enough, all fluid
particles will be mixed by diffusion.

We conclude, therefore, that $\alpha_\text{diff}$ is a more relevant
measure of mixing quality than $\Delta\mathcal{R}$, since the latter is not
always correlated to the fraction of well-mixed particles.

\subsection{Eigenmode index}

We now present a final index of mixing quality based directly on
statistics of the invariant concentration field $\CSE$ of the strange
eigenmode. We show that the normalized standard deviation of $\CSE$ is
a relevant measure of mixing quality, and we examine how this mixing
index depends on the diffusivity of dye.

\subsubsection{Definition}

We define a new index of mixing called \emph{eigenmode index}. The
eigenmode index is defined as the standard
deviation~$\sigma(\CSE)$ of the strange eigenmode divided by its mean
concentration~$\langle \CSE \rangle$:
\begin{equation} 
\sigSE = {\sigma(\CSE)}/{\langle \CSE \rangle}. 
\label{eq:defsigSE}
\end{equation} 
$\sigSE$ is easily measured in decay experiments of type A protocols
as the normalized standard deviation of the concentration field after
a few mixing periods. This quantity becomes time-independent once the
concentration field has achieved an eigenmode pattern, and is
characteristic of the flow for a given value of
diffusivity~\cite{Gouillart2009a}.

$\sigSE$ is a measure of the fluctuations of the strange eigenmode, in
particular of the coverage of the fluid domain by the support of the
eigenmode. The standard deviation of $\CSE$ will be much
greater for a few isolated filaments than for a more uniform coverage
of space by the eigenmode. Small values of $\sigSE$ therefore correspond
to good mixing. If the strange eigenmode covers a fraction
$\Acal$ of the fluid domain, $\sigSE$ can be evaluated to a first
approximation as
\begin{equation} 
\sigSE \simeq \sqrt{\frac{1-\Acal}{\Acal}}\,, 
\label{eq:sigA}
\end{equation}
which is the value $\sigSE$ would take for a constant value of $\CSE$
over the support of the eigenmode. 

\subsubsection{Dependence of the eigenmode index on the diffusivity}

An important property of the eigenmode index $\sigSE$ is its specific
dependence on the diffusion coefficient, or equivalently, the pixel size
at which the concentration field is probed. This dependence is directly
related to the multifractal properties~\cite{Halsey1986} of the unstable
manifold of the chaotic saddle~\cite{Jung1993, Sommerer1996, Tel2000,
Tel2005}, which is the set that supports the strange eigenmode pattern as
we saw in Sec.~\ref{sec:history}.

\begin{figure}
\subfigure[]{
\begin{minipage}{0.46\columnwidth}
{\includegraphics[width=0.98\textwidth]{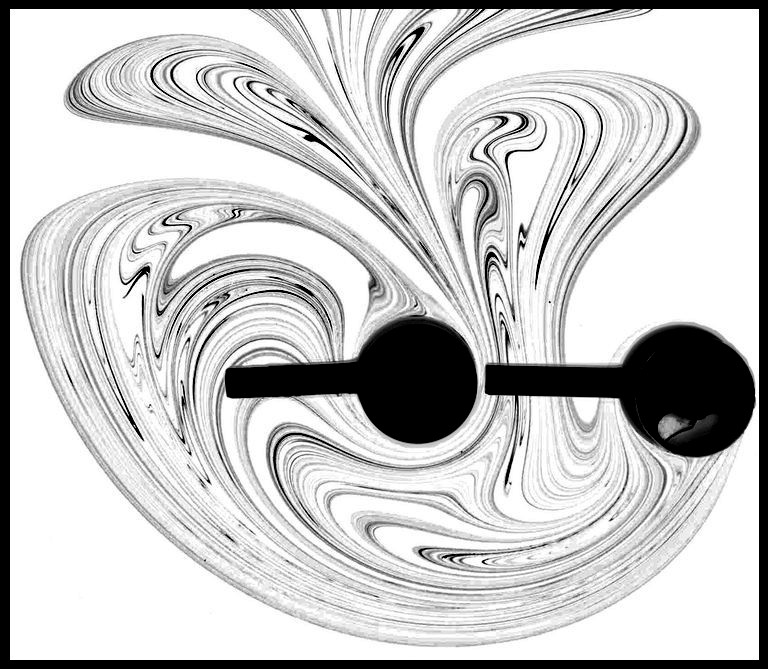}}
\end{minipage}
}
\subfigure[]{
\begin{minipage}{0.46\columnwidth}
{\includegraphics[width=0.98\textwidth]{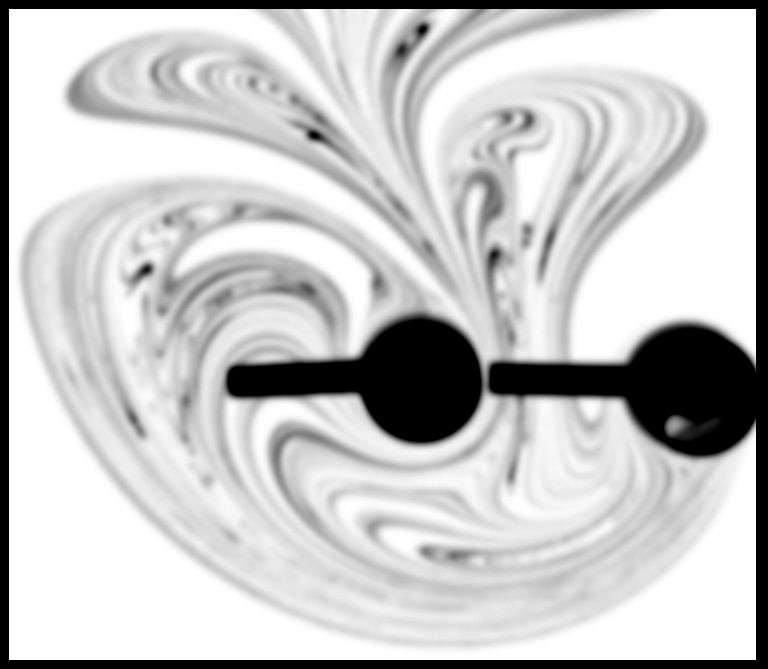}}
\end{minipage}
}
\subfigure[]{
\centerline{\includegraphics[width=0.98\columnwidth]{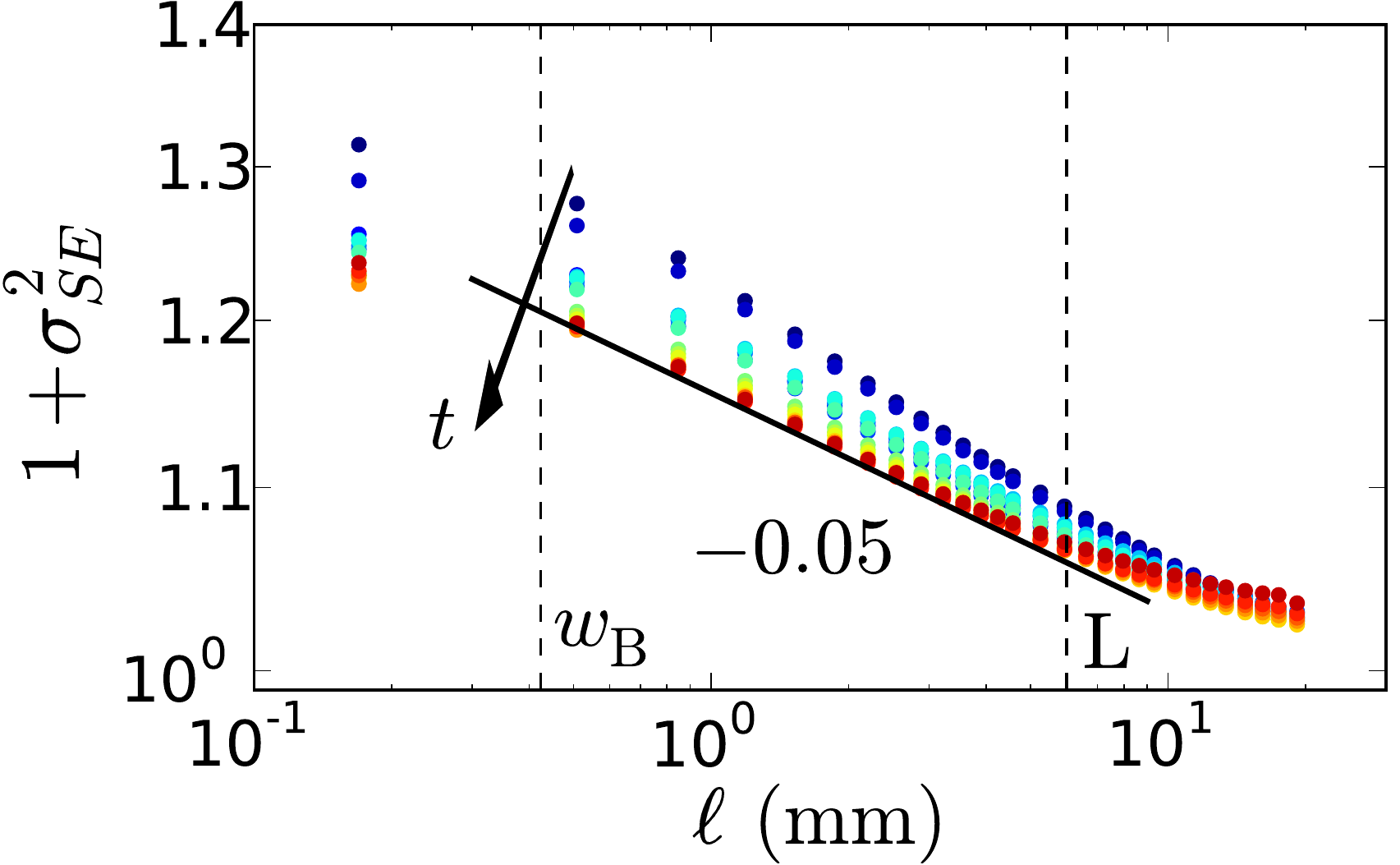}}
}
\caption{(a) Eigenmode pattern observed in a type A protocol. (b) Same
pattern blurred at a scale $\boxw=25$ pixels. (c) Evolution of $1 +
\sigSE^2(t)$ with the blurring scale $\boxw$ plotted in log-log
coordinates, where $\sigSE^2(t)$ is the rescaled variance of the dye
pattern at time $t$. A power-law scaling is observed. According to Eq.
(\ref{eq:fractal_powerlaw}), the small value of the slope ($0.05$) implies
that the correlation dimension $D(2)$ is close to the dimension of space
$d_0$, meaning that the pattern covers the space well. The lower cut-off
at the diffusion scale $\wb$, as well as the upper cut-off imposed by the
size of the largest lobes $L$, are shown as dotted
lines.\label{fig:fractal_manip}}
\end{figure}

In our experiments, the temporal development of the mixing pattern
resembles the construction of an (inhomogeneous) 1D Cantor
set~\cite{Halsey1986}.  The mixing region is stretched and folded, and
a fraction is removed and replaced by dye-free fluid. This is
analogous to the fractal Cantor construction where one repeatedly
removes a fraction of each segment -- the equivalent of an iteration
of the construction step being one stirring period. The fractal
character of the chaotic saddle and its manifolds has been studied
previously~\cite{Jung1993, Sommerer1996, Tel2000, Tel2005}. Studies of
fractal patterns due to chaotic advection have been performed in
closed flows as well, for the classical closed-flow baker's
map~\cite{Ott1989, Antonsen1991}, the blinking vortex
flow~\cite{Fung1991, Muzzio1992}, or the flow between eccentric
cylinders~\cite{Muzzio1992}. Here, we are interested in the
implications of this fractal character for the eigenmode index.

Let us consider the evolution of $\sigSE$ when the pattern is
blurred at different scales. We perform the blurring by convolving the
pattern by a constant step function of width $\boxw$ (in 1D for the
map and in 2D for experimental data). This mimics the action of
a diffusion coefficients $\kappa$ that smoothens the
pattern at the diffusive scale $\boxw = \wb(\kappa) =
\sqrt{\kappa/\la}$ where $\la$ is the mean stretching
(Lyapunov exponent). We have represented in
Fig.~\ref{fig:fractal_manip}(a) and (b) the pattern of the strange
eigenmode for a protocol of type A, and the same pattern blurred by
a square kernel of size 25 pixels (the size of the initial picture in
Fig.~\ref{fig:fractal_manip}(a) is $1700\times1500$ pixels). The
blurring removes the finer details of the pattern by merging together
neighboring strips. It follows from the fractal character of the
pattern that a part of the blurred pattern of size $\delta$ is
statistically equivalent to a part of the initial pattern of size
$\delta/\boxw$. This property implies the following power-law scaling, that we
derive in Appendix~\ref{sec:appendix}:
\begin{equation}
1 + \sigSE^2 \sim \boxw^{D(2) - d_0},
\label{eq:fractal_powerlaw}
\end{equation}
with~$d_0$ the dimension of space and $D(2)$ the fractal
correlation dimension of the support of the eigenmode. The quantity~$d_0 -
D(2)$ is a fractal \emph{codimension} that measures the fraction of
space that the eigenmode fails to cover.

We check the validity of the power-law scaling
\eqref{eq:fractal_powerlaw} for the eigenmode pattern of
Fig.~\ref{fig:fractal_manip}(a). We blur the mixing pattern at different
scales (see Fig.~\ref{fig:fractal_manip}(b)), and compute the resulting
intensity of segregation $\sigSE$. For the domain over which $\sigSE$ is
computed, we select only the filamentary pattern and not the large white
strips close to the channel walls which do not participate in the
stretching and folding process. In addition, we have considered the
concentration field inside the mixing region rather than in the
downstream lobes, as the pattern is larger and has a wider range of
scales so that we expect to see the fractal scaling on a larger range.
The values of $1 + \sigSE^2$ are plotted against the blurring scale
$\boxw$ in Fig.~\ref{fig:fractal_manip} for different mixing times. We
observe a power-law scaling for all curves over a full 1.5 decades in
space, which is the signature of the ongoing stretching and folding
process. All curves converge rapidly onto a single curve as a result of
the convergence to a persistent pattern. At early times the effective
correlation dimension is smaller than the value for the persistent
eigenmode, since at the beginning of an experiment we introduce a small
spot of dye and the coverage of space by the dye is less important than
for the final pattern.

The lower limit of the fractal scaling range corresponds to the
cut-off imposed by the physical molecular diffusion (as opposed to the
artificial blurring we perform) in the experiment. In our setup, the
spatial resolution of the camera is smaller than the diffusive scale
so that blurring the pattern below the diffusive scale does not change
it.  For large values of $\boxw$, no fractal scaling is expected once
the blurring scale is larger than the width of the largest hole in the
pattern.  We verify that this scale corresponds roughly to the end of
the power-law scaling range.  The scaling of~$1 + \sigSE^2$
allows one to extrapolate the value of the mixing index to different
values of the diffusivity.  This is of practical interest for
applications where the substance to be mixed has a different
diffusivity from dye used in a model experiment, or when dealing with
effective numerical diffusivity in a numerical simulation.

\section{Application to continuous injection of dye\label{sec:Is}}

We now turn to the investigation of open-flow mixing for the case of the
continuous injection of dye.

\subsection{The open-flow baker's map}
\label{sec:openbaker}

\begin{figure}
\centerline{\includegraphics[width=0.6\columnwidth]{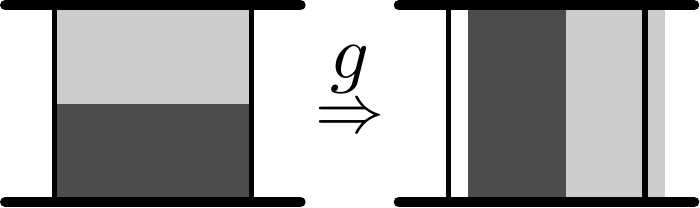}}
\caption{The action of the open-flow baker's map: $\BM$ cuts the
  central square of a channel strip into two strips, stretches them,
  and re-stacks them inside the square shifted downstream to mimic
  advection in a channel. \label{fig:open_map}}
\end{figure}

Let us first introduce an idealized 1D map that will be examined in
parallel with the experiments. The map was described in previous
studies~\cite{Neufeld1998,Gouillart2009a}, and serves as a toy model
for chaotic advection in open flows. It is an open-flow variant of the
well-studied area-preserving baker's
map~\cite{Farmer1983,Ott1989,Antonsen1991,Fereday2002}. The action of
the map is depicted in Fig.~\ref{fig:open_map}. The map is defined
on a channel-like domain, that is an infinite strip composed of a
square central mixing region, and corresponding upstream and
downstream regions.  At each iteration, the action of the map consists
of two parts: (i) a translation of the strip by a distance $\U$,
mimicking the global advection in an open flow; and (ii) a
`traditional' baker's map inside the central mixing region.  The
traditional baker's map consists of a division into two strips, which
are then compressed and re-stacked, preserving area. The map therefore
reproduces the main ingredients of chaotic advection in open flows:
stretching, folding, and escape.

The map has the property of mapping a distribution invariant along the
direction transverse to the channel into another such distribution.
We restrict to such invariant distributions for the upstream
concentration field, so that the action of the map amounts to a 1D
transformation~$\BM$.
Specifically, we study a family of maps defined inside the unit
interval by
\begin{subequations}
\begin{equation}
\BM: x \mapsto \BM_1(x) \cup \BM_2(x),\qquad 0\le x\le 1,
\label{eq:map_baker_a}
\end{equation}
where 
\begin{equation}
\BM_1(x) = \U + \gamma x\,;\qquad \BM_2(x) = \U + \gamma + (1-\gamma)x
\end{equation}
\label{eq:map_baker}%
\end{subequations}%
and the union ($\cup$) symbol in~\eqref{eq:map_baker_a} means
that~$\BM$ is one-to-two: every point~$x$ has two images given
by~$\BM_1(x)$ and~$\BM_2(x)$. The parameter $0<\gamma<1$ controls the
inhomogeneity of stretching; $\gamma=1/2$ corresponding to homogeneous
stretching. Diffusion is mimicked by letting the concentration evolve
according to the heat equation with diffusivity $\kappa$ during a unit time
interval in between iterations of the maps.

Such baker's maps are meant to mimic type A protocols; they cannot
describe type B protocols fully because of the absence of stagnation
points.  As was observed in the experiments, a spot of dye is rapidly
transformed by the map into a persistent pattern, that is, the strange
eigenmode of the map.

\subsection{Case of continuous injection of dye}

\begin{figure}
\centerline{\includegraphics[width=0.97\columnwidth]{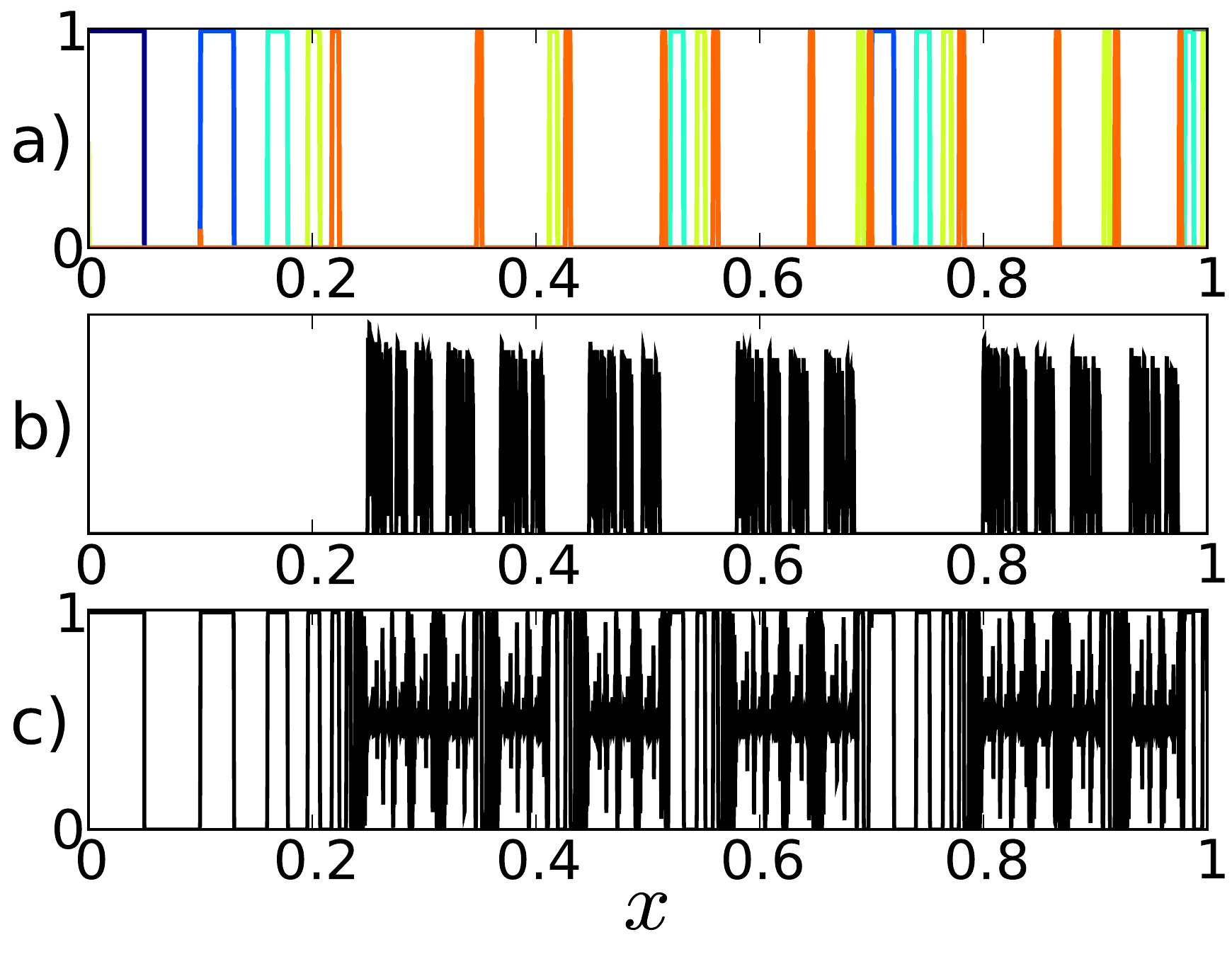}}
\caption{Mixing patterns obtained for the open-flow baker's map: (a)
  For a single square-shaped blob injected at $t=0$ and short
  residence times (plotted with different colors, from blue to red);
  (b) Same initial data, but for a large number of iterations of the
  map, so that the concentration pattern approximates the strange
  eigenmode.  (c) Concentration for a continuous pulse profile
  entering the mixing region. Large fluctuations are obtained in the
  same regions as in (a), that is, for short residence times, while
  fluctuations are much weaker on the support of the strange eigenmode
  visible in (b). \label{fig:scenar_map}}
\end{figure}

For a large class of industrial processes, fluid advected upstream of the
mixing region is always inhomogeneous, and its concentration field
measured in a sufficiently large region is characterized by stationary
statistics. For example, one could realize a mixing experiment by filling
one half of the upstream channel with blue dye and the other half with red
dye, and try to mix together the blue and red streams inside the mixing
region. 

In our experiments, we focused instead on the time-localized injection
of a single spot of dye. Practical reasons for this choice are
twofold. First, the experimental realization of a stationary injection
of dye is more problematic, especially if it has to be carried out
until the downstream concentration field exhibits stationary
statistics.  Second, the choice of an upstream stationary field is
always arbitrary: we may separate the channel into two blue and red
halves, but we might also choose instead a zebra-like pattern of
arbitrary wavelength.  It is likely that any measure of mixing
performed on such experiments will depend on some characteristics of
the upstream pattern, for instance the typical lengthscale of
homogeneity: a thin ``premixed'' zebra pattern will result in a more
homogeneous downstream field than two parallel streams. In contrast,
the strange eigenmode observed for a dye spot can be regarded as a
\emph{unique} signature of the underlying flow.  We will now explain
the relationship between the pattern for stationary injection of dye
and the strange eigenmode resulting from a dye spot.

In the case of a stationary injection of dye, the incoming fluid can
be divided in distinct patches that enter the mixing region at
different times, such as the patch containing the dye spot that we
considered in Sec.~\ref{sec:history}. Inside the mixing region, some
of these patches have been compressed onto the support of the strange
eigenmode, where they are blurred with other patches by diffusion. The
remainder of the mixing region is covered by patches that have not yet
been mixed with others and bear concentration levels similar to the
upstream levels.  Therefore, high fluctuations of the concentration
pattern are found on the complement of the strange eigenmode, whereas
fluid is homogeneous on the strange eigenmode itself. This is
illustrated with the open-flow map in Fig.~\ref{fig:scenar_map}:
fluctuations of the continuous pattern are found in the holes of the
strange eigenmode.

\subsection{Intensity of segregation}

We now describe a measure of mixing commonly used in chemical
engineering for continuous mixing, the \emph{intensity of
  segregation}, and we examine the relation between this quantity and
the eigenmode index.

\subsubsection{Definition}

For a continuous injection of inhomogeneous fluid, the quality of mixing
can be measured by the \emph{intensity of segregation}
\begin{equation}
\sigC = {\sigma_\text{out}}/{\sigma_\text{in}}\,,
\end{equation}
where $\sigma_\text{in}$ and $\sigma_\text{out}$ are respectively the
standard deviation of the concentration field measured on an upstream
(resp. downstream) region chosen to be large enough so that $\sigC$ is
time-independent. $\sigC=0$ corresponds to perfect mixing, and
$\sigC\leq 1$ since mixing can only reduce fluctuations. This quantity was
introduced by Danckwerts~\cite{Danckwerts1952} and is widely used in
chemical engineering.  The quantity $1-\sigC$ can be regarded as the
fraction of fluctuations suppressed in the mixing region. $\sigC$ is
independent of the global intensity of scalar concentration, however it
may depend on the spatial distribution of heterogeneity in the upstream
pattern.

\subsubsection{Link between $\sigC$ and $\sigSE$}

For a poor mixer with a meager coverage of space by the eigenmode,
there are typically large holes between parts of the eigenmode where
the greatest fluctuations of the continuous pattern are nested. In
these holes, concentration levels are of the same order as in the
upstream pattern, since the pattern has been compressed only to a small
extent and has not yet been smeared by diffusion. In such a case the
value of $\sigC$ is expected to be close to $1-\Acal$, the fraction of
space occupied by the complement of the
eigenmode. Using~\eqref{eq:sigA}, we deduce the approximate relation
between $\sigC$ and $\sigSE$ in the limit of poor mixing:
\begin{equation}
\sigSE \simeq \sqrt{\frac{\sigC}{1-\sigC}} \;\;\text{for}\;\; \sigC
\rightarrow 1.
\label{eq:sigsig_poor}
\end{equation}

\begin{figure}
\centerline{\includegraphics[width=0.98\columnwidth]{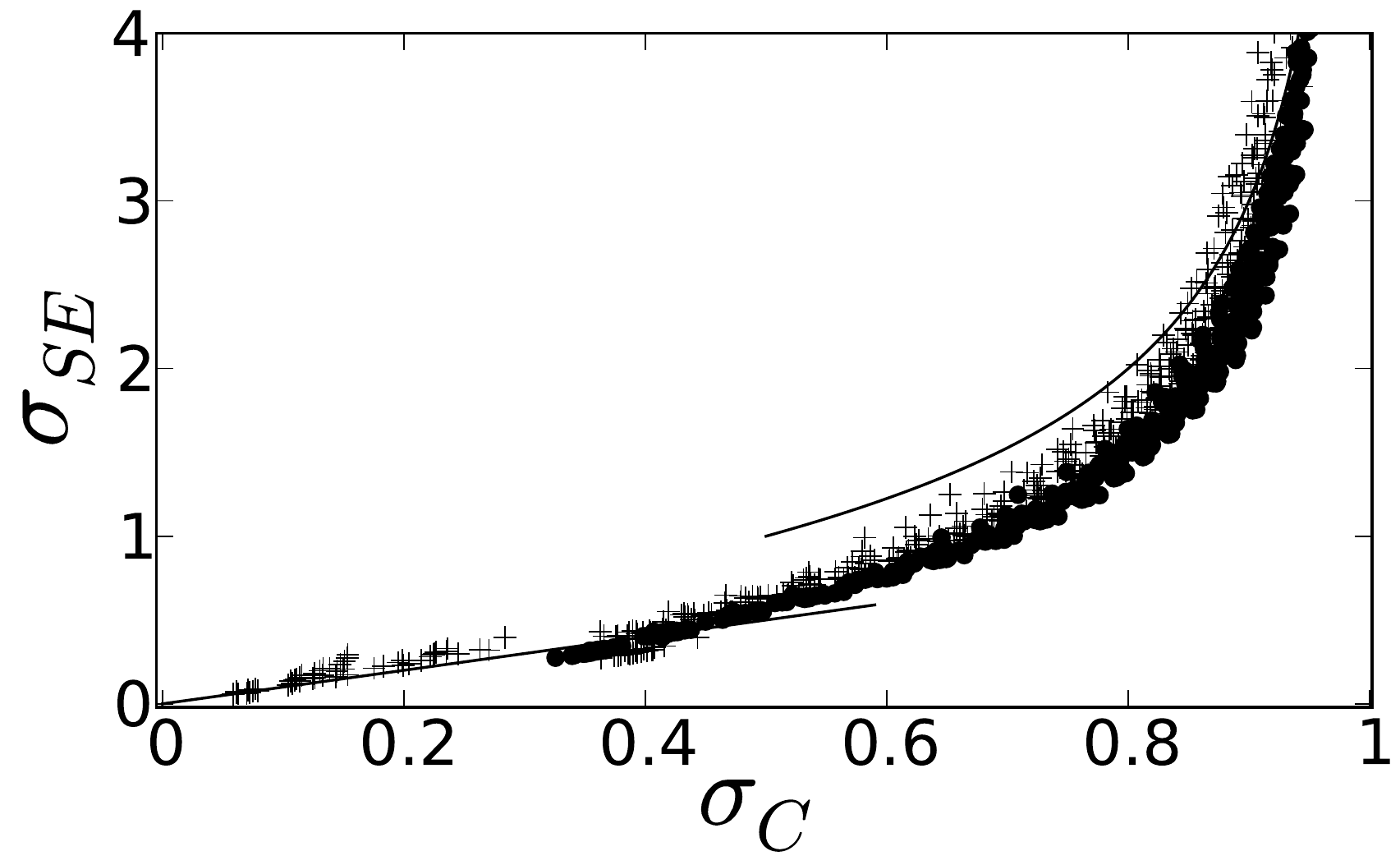}}
\caption{$\sigSE$ vs.\ $\sigC$ for the baker's
  map~\eqref{eq:map_baker} for different values of the parameters $\U$
(ranging from $4\times 10^{-3}$ to 0.2)
  and $\gamma$ (ranging from 0.3 to 0.7). $+$ and $\bullet$
  symbols correspond to two different
  typical scales of the inlet pattern (0.4 and 1). A correlation is
  observed between $\sigSE$ and $\sigC$, becoming linear when they are
  small.  The analytical relations~\eqref{eq:sigsig_poor}
  and~\eqref{eq:sigsig_good} (for poor and good mixing) are plotted as
  solid lines. \label{fig:varphipsi}}
\end{figure}

We now use the open-flow baker's map~\eqref{eq:map_baker} to investigate
the strength of the correlation between $\sigSE$ and $\sigC$. The map
allows us to simulate both a continuous or limited injection of scalar,
and to also vary the parameters $\U$ and $\gamma$ to test different
mixing flows. For the continuous injection of scalar we choose a square
wave of spatial period $\lambda$. $\sigC$ is measured downstream over a
domain of width $100\lambda$. We compute both $\sigC$ and $\sigSE$ for
400 different maps~\eqref{eq:map_baker} where $\U$ and $\gamma$ are
uniformly distributed in the intervals~$[0.02, 0.2]$ and~$[0.3, 0.7]$,
respectively. The comparison is done for two different inlet
wavelengths $\lambda = 0.4$ and $1$. The diffusivity is set to $\kappa =
10^{-8}$, which amounts to setting the diffusive cut-off scale $\boxw
\sim 10^{-4}$. It is apparent in Fig.~\ref{fig:varphipsi} that there is a
strong correlation between the two mixing indices, as expected from the
above qualitative arguments. The dispersion around the mean curve is
quite small and decreases when the wavelength of the inlet profile is
increased. We found the root-mean-square deviation around the mean curve
to be $8\%$ for $\lambda=0.4$, and $5\%$ for $\lambda=1$. In comparison,
we obtain a dispersion of the order of $1\%$ when plotting the intensity
of segregation measured for the two values of the wavelength against each
other. We find the master curve $\sigSE = \flambda(\sigC)$ to depend
slightly on the wavelength of the continuous profile $\lambda$, since the
intensity of segregation depends on $\lambda$. Note that not all the
results of our simulations of the map are plotted in
Fig.~\ref{fig:varphipsi}, as for the largest values of $\U$, and inhomogeneous
values of stretching ($\gamma$ far from $1/2$), the strange eigenmode
scarcely covers the unit interval and values of $\sigSE$ lie outside the
range represented here.

The evident relationship between $\sigSE$ and $\sigC$ suggests that
$\sigSE$ could be used as a convenient substitute for the intensity of
segregation $\sigC$ to evaluate mixing efficiency. Additionally, the
relation between the two measures can be easily understood for the two
extreme regimes of very good or extremely poor mixing. Protocols with
a short mean residence time (e.g., large values of $\U$ for the
map~\eqref{eq:map_baker}) are characterized by eigenmodes that cover a
small fraction of space. In this limit of poor mixing, we expect
$\sigSE$ and $\sigC$ to obey Eq.~\eqref{eq:sigsig_poor}. The
relation~\eqref{eq:sigsig_poor} is plotted in Fig.~\ref{fig:varphipsi}
for comparison with the data. It approximates the data well for $\sigC
\gtrsim 0.8$, that is, for poor mixing.

More interesting for practical purposes is the case of good mixing and
long mean residence times (e.g., small values of $\U$ for the map),
for which it is important to accurately rate protocols according to
their mixing efficiency. The relationship between $\sigSE$ and $\sigC$
can be computed for the special case of an incoming square wave, which
has been used in the simulations.  In the limit of large residence
times, many time-periods are needed before a full spatial
half-wavelength of the inlet pattern has entered the mixing region.
Hence, the color of injected fluid (i.e., black or white) is
homogeneous for long times, as in decay experiments.  The
concentration pattern is close to the strange eigenmode for almost all
times, except during the transition periods of the incoming fluid
between dyed and undyed.  Inside the mixing region, the standard
deviation is therefore proportional to the mean concentration:
\[
\sigma(C, t) = \alpha\, \langle C \rangle (t). 
\]  
By definition~\eqref{eq:defsigSE}, the constant of
proportionality~$\alpha$ equals $\sigSE$.  The quantity $\langle C
\rangle (t)$ oscillates around the mean value of the upstream
concentration on exponential branches corresponding to the decay of
the eigenmode for each color.  Simple algebra then leads to the
relation
\begin{equation}
\sigSE \simeq \frac{\sigC}{\sigma_0} \;\; \text{for small } \sigC,
\label{eq:sigsig_good}
\end{equation}
where $\sigma_0$ is the standard deviation of the upstream pattern
normalized by the mean concentration ($\sigma_0 = 1$ for the square
waves considered here). The above relation is verified on
Fig.~\ref{fig:varphipsi} for small values of $\sigC$.

\section{Conclusions and discussion \label{sec:concl}}

In this article, we have examined quantitative criteria to rank open
flows with chaotic advection according to their mixing efficiency.
Measurements performed for a continuous injection of dye depend strongly on
the spatial distribution of the incoming dye pattern, thus obscuring the
inherent efficiency of the mixing device.  To overcome this difficulty,
we have proposed a measure of mixing based on the strange eigenmode of
the flow that depends only on the mixing flow and the scalar diffusivity.
Our \emph{eigenmode index} is defined as the standard deviation $\sigSE$
of the eigenmode and measures approximately the area covered by fluid
particles that are well mixed.

As expected, we found a good correlation between $\sigSE$ and measures
of the variance for the continuous injection of a periodic pattern
with a fixed lengthscale.  The measure~$\sigSE$ therefore ranks mixing
protocols in the same order as Danckwerts' intensity of
segregation. The reason for this correlation is that fluctuations
corresponding to short residence times inside the mixing region are
found on the complementary subspace of the strange eigenmode. We have
also noted that the dependence of~$\sigSE$ on the scalar diffusivity
can be predicted if the resolution of the eigenmode pattern is fine
enough to obtain its fractal correlation dimension.  As $\sigSE$ is
easy to measure and has physical and practical relevance, we advocate
the use of this new criterion for evaluating mixing efficiency.

At first sight, $\sigSE$ may seem of little use for protocols (such as
our type B protocol) where a true strange eigenmode is not present
because of regions of slow stretching that trap fluid particles for long
times~\cite{Gouillart2009a}. The vicinity of ``sticky'' elliptical
islands or no-slip solid walls are examples of regions that will be
covered very slowly by an incoming spot of dye.  Conversely, dye will be
trapped in such regions for longer times than in the rest in the mixing
region, so that no persistent eigenmode is
observed~\cite{Gouillart2009a}. However, these regions contribute only
weakly to fluctuations in the outflowing fluid for the case of a
continuous injection of dye, since particles that finally escape into the
rest of the chaotic region have typically experienced much more
stretching than fluid particles with short residence times. Therefore, we
suggest measuring $\sigSE$ at intermediate times when a quasi-persistent
pattern is observed, that is, when dye filaments have reached the
Batchelor scale everywhere but on these small regions of anomalous
stretching. Such a measure will slightly underestimate the efficiency of
mixing as compared to other protocols, because some areas not yet covered
by dye will contribute to $\sigSE$, whereas they do not contribute to
fluctuations of the continuous pattern. Nevertheless, this is only a
small effect. $\sigSE$ should not be measured at long times when dye is
mostly concentrated in the slow-stretching regions, since the
fluctuations of this pattern cannot be related to those for continuous
injection.  Most efficient mixing protocols display either a true strange
eigenmode, or a ``quasi'' strange eigenmode corresponding to a
quasi-permanent pattern observed during a large range of intermediate
times.  Otherwise, a large portion of the mixing region is
occupied by fluid that is trapped for long times. This may happen, for
example, if the rods do not come close enough to the channel walls for a
type B protocol. As the mean residence time in the mixing region is
determined solely by the flowrate, this implies that there will be an
increased fraction of short residence times, hence bad mixing.

Finally, we note that our study of mixing efficiency was confined to a
global characterization of the homogeneity of the outgoing fluid. In
particular, no characterization of the \emph{spatial} fluctuations of
the dye pattern has yet been proposed for open flows. It may be
relevant in practical applications to measure the distribution of
widths of strips of dye exiting the mixing region. Once again, this
measure could be replaced by the simpler characterization of the
widths of holes in the strange eigenmode, where the largest
fluctuations are found. This would require the development of
dedicated image processing tools.


\section*{Acknowledgements}

We are grateful to Franck Pigeonneau and St\'ephane Roux for
enlightening discussions, to Beno\^it Roche for help with image
processing, and to C\'ecile Gasquet and Vincent Padilla for technical
assistance.  J-LT was supported by the US National Science Foundation
under grant DMS-0806821.


\appendix
\section{Scaling of $\sigSE$ with the diffusivity or box size
\label{sec:appendix}}

Let us derive here the evolution of $\sigSE$ with the box size $\boxw$,
using the multifractal properties of the support of the strange eigenmode.

The multifractal spectrum of a fractal set is defined as
follows~\cite{Halsey1986}. Let $p_\mathcal{B}$ be the measure in the box
$\mathcal{B}$ of
size $\boxw$, and $q \in \mathbb{R}$. The invariance of the set at different scales implies the
following scaling of the partition function $Z(q, \boxw)$:
\begin{equation}
Z(q, \boxw) = \sum_\mathcal{B} p_\mathcal{B}^q \sim \boxw^{\tau(q)}\,. 
\end{equation} 
The spectrum of multifractal dimensions is obtained from the spectrum
$\tau(q)$ as
\[
D_q = \tau(q)/(q-1).
\]

 Using the definition of $\sigSE$, we write
\begin{equation}
1 + \sigSE^2 = \frac{\sum_i C_i^2}{\bigl(\sum_i C_i\bigr)^2}\,, 
\end{equation}
with $C_i = \C(\xb_i)$ the concentration level in pixel~$i$.  To find the
multifractal scaling, instead of summing over pixels we sum the
concentration over larger square boxes $\mathcal{B}$ of size $\boxw$,
over which the concentration has been blurred and is nearly a constant
$\tilde{C}(\mathcal{B})$.
Since
\[
\sum_{i\in\mathcal{B}} C_i^q = \boxw^{d_0} \tilde{C}^q(\mathcal{B})
\]
with~$d_0$ the dimension of
space and $q \in \mathbb{R}$, we obtain
\begin{equation*}
1 + \sigSE^2 = \frac{1}{\boxw^{d_0}} \sum_{\mathcal{B}}
\left(\frac{\tilde{C}}{\sum_\mathcal{B} \tilde{C}} \right)^2
= \frac{1}{\boxw^{d_0}} \sum_{\mathcal{B}} p^2_\mathcal{B},
\end{equation*}
with the measure
$p_\mathcal{B}=\tilde{C}/\sum_\mathcal{B} \tilde{C}$. Using the
definition of the multifractal spectrum, we finally get
\begin{equation}
1 + \sigSE^2 \sim \boxw^{D(2) - d_0}.
\end{equation}

\end{document}